\documentclass[preprint,pre,superscriptaddress,showpacs]{revtex4}

\usepackage{graphicx}% Include figure files
\usepackage{dcolumn}% Align table columns on decimal point
\usepackage{bm}% bold math
\newcommand*\Bell{\ensuremath{\boldsymbol\ell}}

\begin{document}

%\preprint{AIP/123-QED}

\title[Random Close Packing from Hard Sphere Percus-Yevick Theory]{Random Close Packing from Hard Sphere Percus-Yevick Theory}

\author{Eytan Katzav}
 \email{eytan.katzav@mail.huji.ac.il}
 \affiliation{Racah Institute of Physics, The Hebrew University, Jerusalem 91904, Israel} 
%\altaffiliation[Also at ]{Physics Department, XYZ University.}

\author{Ruslan Berdichevsky}
 \email{rusalmighty@gmail.com}%
\author{Moshe Schwartz}
 \email{bricki@netvision.net.il}
 \affiliation{Department of Physics, Raymond and Beverly Sackler Faculty of Exact Sciences,\\ Tel Aviv University, Tel Aviv 69978, Israel}%

% \homepage{http://www.Second.institution.edu/~Charlie.Author.}
%\affiliation{%
%Second institution and/or address%\\This line break forced% with \\
%}%

\date{\today}% It is always \today, today,

\begin{abstract}
The Percus-Yevick theory for monodisperse hard spheres gives very good results for the pressure and structure factor of the system in a whole range of densities that lie within the liquid phase. However, the equation seems to lead to a very unacceptable result beyond that region. Namely, the Percus-Yevick theory predicts a smooth behavior of the pressure that diverges only when the volume fraction $\eta$ approaches unity. Thus, within the theory there seems to be no indication for the termination of the liquid phase and the transition to a solid or to a glass. In the present article we study the Percus-Yevick hard sphere pair distribution function, $g_2(r)$, for various spatial dimensions. We find that beyond a certain critical volume fraction $\eta_c$, the pair distribution function, $g_2(r)$, which should be positive definite, becomes negative at some distances. We also present an intriguing observation that the critical $\eta_c$ values we find are consistent with volume fractions where onsets of random close packing (or maximally random jammed states) are reported in the literature for various dimensions. That observation is supported by an intuitive argument. 
This work may have important implications for other systems for which a Percus-Yevick theory exists.
\end{abstract}

\maketitle

\section{Introduction}

The hard sphere model provides a canonical minimalistic model that captures the main ingredient in the description of simple liquids, namely the strong short-range repulsion between atoms in the liquid. As in other systems in equilibrium statistical physics, the model is used to obtain macroscopic observables from the microscopic description of the system. In the case of the hard sphere model the goals are the equation of state, the liquid structure factor and a description of the solidification of the liquid in terms of the average particle density, $\bar \rho $ and $R$, the range of the hard sphere interaction (namely the diameter of the hard spheres). A wide arsenal of methods has been applied over the years to the hard sphere problem with considerable success. Monte-Carlo and Molecular Dynamics simulations have been applied to that model as early as the 1950s \cite{Wood1957,Alder1957,Alder1959} and extended much later. 
Experimental studies, first with ball-bearings \cite{Bernal1960,Bernal1964} and later with colloids \cite{Vladeeren1995}, helped elucidate properties of very dense packings which where not attainable in simulations. This gave rise to the concept of the random close packing (RCP) density, defined as the maximal density among amorphous packings. Improved computation technologies went hand in hand with hard-sphere simulations. For example, the hard sphere system is one of the first systems to be simulated on the Small Web Computing (SWC) platform in recent years \cite{Bishop2005}. These important numerical efforts resulted in obtaining the phase diagram of the system, including crystallization and a super dense rotation invariant phase in three dimensions \cite{Hoover1968,Frenkel1984,Bolhuis1996,Gruhn2001}. The most trusted analytic tool applied successfully to the hard sphere problem is the virial expansion, which is based in turn on the cluster expansion \cite{Mayer1941,Morita1961,Dominicis1962,Dominicis1963,Stell1964,Hansen2006,Clisby2005,Clisby2006,Zhang2014}.

The other two interesting analytic approaches are the Hyper-Netted-Chain (HNC) approximation \cite{Rowlinson1965} and the Percus-Yevick (PY) equation \cite{Percus1958} for the structure factor of the hard sphere system. The most appealing to us is the PY equation and that is for several reasons. First the equation has been given exact analytic solutions in odd dimensions $d \le 7$ \cite{Thiele1963,Wertheim1963,Wertheim1964,Baxter1968a,Leutheusser1984,Rohrmann2007,Robles2007,Robles2004} (where $d$ is the dimension of the system). In fact, an exact analytic solution can be obtained in principle for any odd dimension but it involves solving a polynomial equation of degree $2^{(d-3)/2}$. Thus, the highest dimension for which a strict analytic solution in closed form exists is $7$, due to the Abel-Ruffini theorem. However, thanks to the existence of this developed analytic structure it is possible to obtain semi-analytic results for higher odd dimensions with a simple numerical computation \cite{Robles2004,Rohrmann2008}. More recently, a systematic analytic method of solution, based on the virial expansion for the PY equation, has been obtained for general dimensions including the even ones \cite{AddaBedia2008a,AddaBedia2008b}.

The paper is organized as follows: In section II we present an alternative derivation of the Percus-Yevick equation. In section III we show that unlike the common lore the PY theory breaks down at a critical density, which is intrinsic to the theory. In section IV we discuss the physical meaning of the reported observations and in section V we summarize and provide future perspective.

\section{An alternative derivation of the Percus-Yevick equation}
The PY equation is usually seen as a certain diagrammatic approximation or closure scheme of the full problem \cite{Hansen2006}. It was shown, however, that the PY equation for the hard sphere system can be given a very simple and intuitive meaning \cite{Schwartz1999,Edwards2003}, which elucidates the assumptions underlying it, as well as possible extensions and refinements. For the benefit of the readers we reproduce here a short derivation of the PY equation. Consider the particle number density, 

\begin{equation}
\rho ({\bf{r}}) = \sum\limits_{i = 1}^N {\delta ({\bf{r}} - {{\bf{r}}_i}}) ,
\label{eq:1}
\end{equation}

\noindent
where ${{\bf{r}}_i}$ is the location of particle $i$ which is one of $N$ identical particles enclosed in a cubic container of linear size $L$ and periodic boundary conditions.
The pair distribution function,
\begin{equation}
g_2({\bf{r}}) = \frac{1}{{\bar \rho }^2} 
\left\langle {\rho ({\bf{0}})\rho ({\bf{r}})
- \bar \rho \delta({\bf{r}})} \right\rangle
\label{eq:2}
\end{equation}

\noindent
yields the $d$ dimensional distribution to find a particle at ${\bf{r}}$ given the existence of another particle at ${\bf{0}}$. The hard sphere system is then viewed as an ideal gas with a pair distribution function which is constrained to vanish for $\left| {\bf{r}} \right| < R$ (where $R$ is the diameter of the hard spheres). To see how it works we have to transform from particle coordinates to collective coordinates \cite{Schwartz2002,Edwards2003} as described shortly in the following.
The natural collective coordinates are the Fourier components of the density,
\begin{equation}
\rho_{\bf{q}} = \frac{1}{\sqrt N}\int {d{\bf{r}}\rho ({\bf{r}}){e^{ - i{\bf{q}} \cdot {\bf{r}}}}} ,
\label{eq:3}
\end{equation}

\noindent
for  ${\bf{q}} \ne 0$  and with discrete Fourier modes ${\bf{q}}_\ell = \frac{2\pi n_\ell }{L}$, where $n_\ell$ is an integer. The ideal gas Fokker-Planck equation for the distribution of the $N$ free particles is translated into a functional Fokker-Planck equation for the probability to obtain a given configuration of the density, $P^I\{ \rho \} $ \cite{Edwards2003}, which reads at equilibrium,
\begin{equation}
\frac{\partial P^I}{\partial t}=
\frac{k_B T}{\sqrt{N}}
\left[ {\sum\limits_{{\bf{k}},\Bell } {\bf{k} \cdot \Bell 
\frac{\partial}{\partial \rho_{\bf{k}}} {\rho_{{\bf{k}} + \Bell }}
\frac{\partial}{\partial \rho_{\Bell}}} 
- \sqrt N \sum\limits_{\bf{k}} {{\bf{k}}^2 
\frac{\partial}{\partial \rho_{\bf{k}}} {\rho_{\bf{k}}}} } \right]
P^I\{\rho \} =0 \, ,
\label{eq:4}
\end{equation}

\noindent
where $\rho_0 = \sqrt N$ is not a dynamical variable and $T$ is the absolute temperature. 

%We are interested in the steady-state measure $P^I_S\{ \rho \}$ and hence $\partial P^I_S / \partial t =0$. Approximating this equation by keeping only the bilinear part in the operators ${\rho _{\bf{k}}}$ and $\frac{\partial }{{\partial {\rho _{\bf{k}}}}}$ for ${\bf{k}} \ne 0$ (this scheme is also known as the Random-Phase-Approximation) we obtain the ideal gas equation
%\begin{equation}
%\sum\limits_{\bf{k}} {{\bf{k}}^2}\frac{\partial }{{\partial {\rho _{\bf{k}}}}}\left[ {\frac{\partial }{{\partial {\rho _{ - {\bf{k}}}}}} + {\rho _{\bf{k}}}} \right]
%P_{S}^I \{\rho\} = 0 .
%\label{eq:5}
%\end{equation}

%\noindent
%This allows for a solution with the following equilibrium distribution, 

%\begin{equation}
%P_{S}^I\{\rho\} \propto \exp \left[ { - \frac{1}{2}\sum\limits_{\bf{k}} {{\rho _{\bf{k}}}{\rho _{ - {\bf{k}}}}} } \right] ,
%\label{eq:6}
%\end{equation}

%\noindent
%which yields the exact result for the structure factor of the ideal gas, $S^I(k) = \left\langle {\rho_{\bf{k}} \rho_{-{\bf{k}}}} \right\rangle  = 1$. (This does not imply that higher order correlations obtained in the above approximation are exact.) 

To make the pair distribution function vanish within the hard sphere range we introduce into Eq. (\ref{eq:4}) a Lagrange multiplier function $\lambda_{\bf{k}}$, which is a Fourier transform of a yet unknown function $\lambda (\bf{r})$, that vanishes outside the hard sphere interaction range. The last requirement reflects the fact that the pair distribution function is constrained only within that range. The equation now reads,

\begin{equation}
\left \{ \sum\limits_{\bf{k}} {{\bf{k}}^2}
\frac{\partial}{\partial \rho_{\bf{k}}}
\left[ {\frac{\partial}{\partial {\rho_{-{\bf{k}}}}}
+ \rho_{\bf{k}} + \lambda_{\bf{k}} \rho_{\bf{k}}} \right] 
-\frac{1}{\sqrt N} \sum\limits_{{\bf{k}},\Bell } 
{\bf{k} \cdot \Bell \frac{\partial}{\partial \rho_{\bf{k}}} {\rho_{{\bf{k}} + \Bell }}
\frac{\partial}{\partial \rho_{\Bell}}} \right\}
P_{S}^I\{\rho\} = 0 .
\label{eq:7}
\end{equation}

\noindent
%Thus, the term $\rho_{\bf{k}}$ in the square brackets on the left hand side of q. (\ref{eq:5}) is replaced by $\left(1 + \lambda_k \right) \rho_{\bf{k}}$. This results in the following structure factor 

\noindent
Note that by making $\lambda$ depend only on the absolute value of $\bf{k}$ ,we assume implicitly a rotation invariant phase. Also, the subscript $S$ on the density distribution function denotes steady state, which is not necessarily the equilibrium state. The reason is that at densities corresponding to the equilibrium solid phase we constrain our system to have a spherically symmetric structure factor. Thus the state described by the solution of Eq. (\ref{eq:7}) and the constraint on the pair distribution function is a steady state metastable state rather than an equilibrium state. Multiplication of Eq. (\ref{eq:7}) by $\frac{1}{2} \rho_k \rho_{-k}$ and functional integration by parts yields the structure factor in terms of the unknown Lagrange multiplier $\lambda_{\bf{k}}$, namely

\begin{equation}
S^{HS}({\bf{k}}) = \frac{1}{1 + \lambda_{\bf{k}}} , 
\label{eq:8}
\end{equation}

\noindent
where $\lambda_{\bf{k}}$ has to obey the two following conditions, 

\begin{equation}
\lambda (r) = \int d{\bf{k}}{\lambda_{\bf{k}}}e^{i{\bf{k}} \cdot {\bf{r}}}  = 0
\quad {\text{for }} r > R ,
\label{eq:9}
\end{equation}
and
\begin{equation}
g_2(r) = 1 + \frac{1}{N} \sum\limits_{{\bf{k}} \ne {\bf{0}}}
\left( S^{HS}({\bf{k}}) - 1 \right){e^{i{\bf{k}} \cdot {\bf{r}}}}  = 0
\quad {\text{for }}r < R .
\label{eq:10}
\end{equation}

\noindent
It turns out that these two conditions are in fact the hard sphere PY equation. In classical liquid theory a quantity termed direct correlation function is used extensively and is traditionally denoted by $c(r)$. In our language, the Lagrange multiplier function $\lambda (r)$ is simply $ - \bar \rho c(r)$ (see Appendix A for more details on the notation used here).

% As can be better appreciated at this point, the PY is a steady-state solution for a rotation invariant (hence disordered) state by construction, for an approximated RPA scheme.  Actually, within this framework the PY equation is just the lowest order theory in the Self-Consistent Expansion of the full model defined by 
%
%\begin{eqnarray}
%&& \left[ \sum\limits_{{\bf{k,\Bell}}} {{\bf{k}} \cdot \Bell \frac{\partial }{{\partial {\rho _{\bf{k}}}}}{\rho _{{\bf{k}} + \Bell}}\frac{\partial }{{\partial {\rho_{\Bell}}}}}  \right. \nonumber \\
%&& \left . - \sqrt N \sum\limits_{\bf{k}} {{\bf{k}}^2}\frac{\partial }{{\partial {\rho _{\bf{k}}}}} (1+\lambda_k)\rho_{\bf{k}} \right]
%P_{S}^{HS}\{\rho\} = 0 .
%\label{eq:11}
%\end{eqnarray}

%\begin{equation}
%\left[ \sum\limits_{{\bf{k,\Bell}}} {{\bf{k}} \cdot \Bell \frac{\partial }{{\partial {\rho _{\bf{k}}}}}{\rho _{{\bf{k}} + \Bell}}\frac{\partial }{{\partial {\rho_{\Bell}}}}} - \sqrt N \sum\limits_{\bf{k}} {{\bf{k}}^2}\frac{\partial }{{\partial {\rho _{\bf{k}}}}} (1+\lambda_k)\rho_{\bf{k}} \right] P_{eq}^{HS}\{\rho\} = 0 .
%\label{eq:11}
%\end{equation}

\noindent
%Note that this equation includes terms tri-linear in the operators $\rho_{\bf{k}}$ and $\frac{\partial}{\partial \rho_{\bf{k}}}$ in addition to the bi-linear terms considered previously. Eq. (\ref{eq:11}) belongs thus 

Although Eq. (\ref{eq:7}) includes terms tri-linear in the operators $\rho_k$ and  $\frac{\partial}{\partial \rho_k}$ , the solution Eq. (\ref{eq:8}) is exact , because of the specific form of those terms. The next step beyond PY can be affected by constraining the triplet  distribution function to vanish whenever any pair of the triplet is closer than the hard sphere interaction range. The equation for the distribution in that case will include tri-linear terms of a different  nature, which will prevent an exact solution for the structure factor in terms of the two point Lagrange multiplier function. The equation in that case belongs, however, to a wide family of stochastic nonlinear systems, described by a functional Fokker-Planck equation that has been treated successfully by the Self-Consistent Expansion (SCE) \cite{Schwartz2008,Schwartz1992,Schwartz1998,Katzav1999,Katzav2003,Katzav2007,Edwards2002}. Thus, the PY equation is not the last word, as it can be systematically improved to include higher order correlation functions. The interesting thing is that in spite of its simplicity, the equation of state the PY approximation produces is in very good agreement with simulations for liquids \cite{Hansen2006}. In fact, since the PY is only an approximation it produces two very good but different equations of state, depending on the route of derivation. When a proper weighted average of the two is constructed the really excellent Carnahan--Starling (CS) equation of state \cite{Carnahan1969} is obtained

\begin{equation}
P_{\rm{CS}} = \bar \rho {k_B}T\frac{1 + \eta  + \eta^2 - \eta ^3}{(1 - \eta )^3} ,
\label{eq:12}
\end{equation}

\noindent
where $P_{\rm{CS}}$ is the pressure and the volume fraction $\eta$ is given (in three dimensions) by

\begin{equation}
\eta  = \frac{\pi R^3}{6}\bar \rho  , 
\label{eq:13}
\end{equation}

\noindent
with similar expressions in other dimensions (see Appendix A). Recall that $R$ is the range of the hard sphere interaction, namely, the diameter, and not the radius of a single sphere.

Since the CS equation of state holds for volume fractions below crystallization, the fact that it holds also above crystallization seems to be irrelevant. The reason is that the PY approximation assumes invariance under rotation and the emergence of a crystalline structure is just due to the fact that the free energy associated with the solid is lower than the one associated with the rotation invariant phase. It is interesting to note, however, that the hard sphere system possesses a metastable super dense rotation invariant phase which is disordered. Actually, the pressure in that phase is well described by the CS equation of state up to $\eta  = 0.57$ for monodisperse hard spheres, while for polydisperse hard spheres the related BMCSL approach \cite{Boublik1970,Mansoori1971} extends way beyond that \cite{Berthier2009,Berthier2016}. This super dense branch should, however, have terminated at random close packing, where the pressure is expected to diverge \cite{Parisi2010}. Furthermore, that branch as predicted by PY continues into nonphysical volume fractions, even above the crystalline close packing. The main trouble with PY is therefore that there seems to be no intrinsic indication within the PY theory that something goes wrong at higher volume fractions. The first message of the present article is that, contrary to the above statements, an intrinsic indication for the failure of the theory at a certain density does exist in PY.

\section{The breakdown of PY at $\eta_c$}

Consider the pair distribution function $g_2({\bf{r}})$ defined above in Eq. (\ref{eq:2}). $g_2 ({\bf{r}})$ is obtained, within the PY approximation, in the following way. The exact solution in odd $d$ dimensions provides the so-called direct correlation function $c(r) \equiv - \lambda (r)/\bar \rho$, for $r < R$. As it happens, those are polynomials of degree $d$ in $r$ with coefficients that are functions of the volume fraction $\eta$. Since for $r > R$, the direct correlation vanishes, obtaining the corresponding Fourier transform $\lambda_k$'s is a straightforward analytic calculation. The last step to obtain $g_2(r)$ is to use Eq. (\ref{eq:8}) for the structure factor and finally use Eq. (\ref{eq:10}) to obtain $g_2(r)$ in the limit of infinite volume by numerical integration. 

\begin{figure}
\includegraphics[width=7cm]{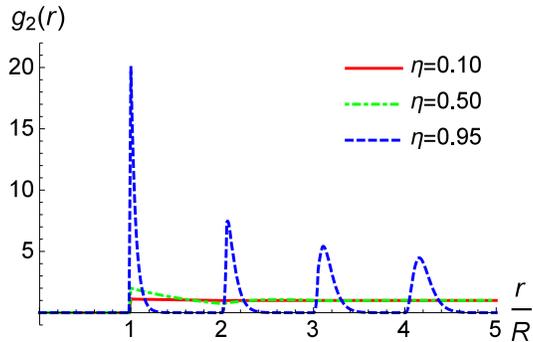}% Here is how to import EPS art
\caption{\label{fig:1} The one dimensional pair distribution functions for low, intermediate and high volume fractions.}
\end{figure}

We begin with the one-dimensional case. In Fig. \ref{fig:1} we present the pair distribution function for three different volume fractions in one dimension. Of particular interest is the high volume fraction graph. The apparent peaks are related to the short range order in the system but it is clear enough that nothing spectacular happens as the peaks are broadened and reduced in height as a function of the distance.

\begin{figure}
\includegraphics[width=7cm]{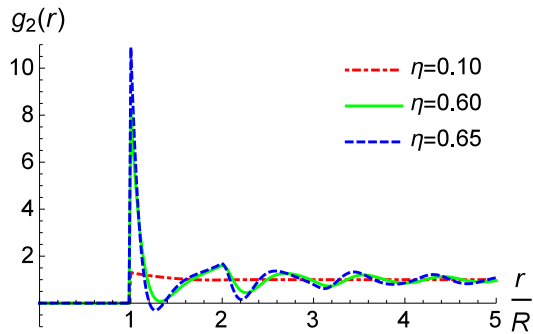}% Here is how to import EPS art
\caption{\label{fig:2} The three dimensional pair distribution functions for low, intermediate and high volume fractions.}
\end{figure}

We continue with Fig. \ref{fig:2}, where we present the corresponding pair distribution function in three dimensions.

\noindent
The low volume fraction graph shows no interesting features, but the intermediate and high density graphs show clearly short range order, which is manifested by the oscillations of the pair distribution function. The alert reader may have already detected a serious problem in the $\eta = 0.65$ case. The pair distribution function, $g_2(r)$ as defined by Eq. (\ref{eq:2}), is by definition nonnegative, while in Fig. \ref{fig:2} the pair distribution function is negative in a certain region of $r$.
Since we also compare our numerical integration with an exact representation of $g_2(r)$ \cite{Smith1970,Kelly2016} which is available in three dimensions up to $r \le 5R$ this negativity cannot be attributed to an artifact of the approximate numerical integration needed to obtain the pair distribution function. More quantitatively, we can look for the lowest volume fraction for which a negative part appears. This is actually the point $r_c$ where both the function and its derivative become zero simultaneously, i.e. $g_2(r_c)=g'_2(r_c)=0$. Using the analytical result of Refs. \cite{Smith1970,Kelly2016} we find that the lowest volume fraction for which a negative part appears is $\eta_c(d = 3) \simeq 0.612574..$.
Similar regions of negative values of the colloid-colloid pair distribution function were obtained for colloids immersed in a fluid \cite{Henderson1986}. 

% as can be seen in Fig. \ref{fig:3}.
%\begin{figure}
%\includegraphics[width=7cm]{PY-3Db}% Here is how to import EPS art
%\caption{\label{fig:3} The three dimensional PY pair distribution function: a zoom-in on the region where it becomes negative.}
%\end{figure}

The inevitable conclusion thus is that the PY approximation breaks down intrinsically at high enough volume fraction. Namely, in contrast to the reasoning based on the continuity of the equation of state across physically impossible densities that was discussed above, the PY calculation itself indicates that something must go wrong, by giving negative values to a function that is nonnegative definite. This is  obviously good news since it sets an internal limit, $\eta_c$, on the applicability of the fluid equation of state (or rotationally invariant phase) at high volume fractions. 
%It is also clear, following, the derivation of the PY equation given in the introductory part of the paper, that the origin of the failure of the PY equation is the Random Phase Approximation, which leads from the exact equation (\ref{eq:11}) to the approximate equation (\ref{eq:7}).

At this point it is natural to ask whether this $\eta_c$ has any physical meaning beyond being an intrinsic upper bound on the theory? The first thing to check is whether it happens in higher dimensions as well. We have obtained the pair distribution function for dimensions $3 < d \le 9$. For the odd dimensions 
we used the exact solution, with a numerical solution of the appropriate polynomial equation when required, as well as a numerical Fourier transform to obtain the pair distribution function - see Refs. \cite{Robles2004,Rohrmann2008} and the appendices for more details (in these dimensions there is not any direct analytical representation of the pair distribution function like the one available in $d=3$ in Refs. \cite{Smith1970,Kelly2016}). It turns out that similar to three dimensions, for all odd dimensions in the range $5 \le d \le 9$, the pair distribution function becomes negative at some range of $r/R$. and above some critical value of the volume fraction, $\eta_c(d)$. The results are summarized in Table \ref{tab:1} below, as well as graphically in Fig. \ref{fig:5} (red circles). The results of our numerical integration are supported by the exact 3D result \cite{Smith1970,Kelly2016} as well as intrinsically by comparison with improved approximate integration.

For even dimensions we use the method and results reported in previous work \cite{AddaBedia2008a,AddaBedia2008b} that provides the pair distribution function as a power series in the volume fraction $\eta$ as

\begin{equation}
g_2(r) = 1 + \sum\limits_{n \ge 1} \eta^n g_2^{(n)}(r) .
\label{eq:14}
\end{equation}

\noindent
In practice, the expansion functions, $g_2^{(n)}(r)$ up to $n = 13$ for $d = 4,6$ and $8$ are available numerically from Ref. \cite{AddaBedia2008b}. These series work very well for small volume fractions. However, in the current work we are interested in fairly high volume fractions and, in particular, in identifying the lowest volume fraction for which $g_2(r)$ develops a negative part. Note that generically $g_2(r)$ is a decreasing function, exhibiting oscillations that become more and more pronounced as the density rises. Based on this observation (and on the odd dimensional cases discussed above) the first negative part should appear at the first minimum of $g_2(r)$, which is obtained in the interval $1 < r/R < 2$. The technical difficulty we encounter is that the radius of convergence of the series Eq. (\ref{eq:14}) is not large, and scales as $2^{-d}$ as the dimension, $d$, grows (see Ref. \cite{AddaBedia2008b} for a more complete discussion). In particular, for the densities that are of interest the series does not converge, and we need to use some method to re-sum it or analytically-continue it. One such popular method is the Pad\'e approximation \cite{Bender1999}. We look at various Pad\'e approximants of $g_2(r)$, which are composed of a polynomial of order $N$ in $\eta$ divided by a polynomial of order $M$ in $\eta$, of the general form 

\begin{equation}
g_2 (r) \simeq \frac{\sum_{n = 0}^N \eta^n u_n(r)}
{\sum_{n = 0}^M \eta^n d_n(r)} \, .
\label{eq:15}
\end{equation}

\begin{figure*}
\includegraphics{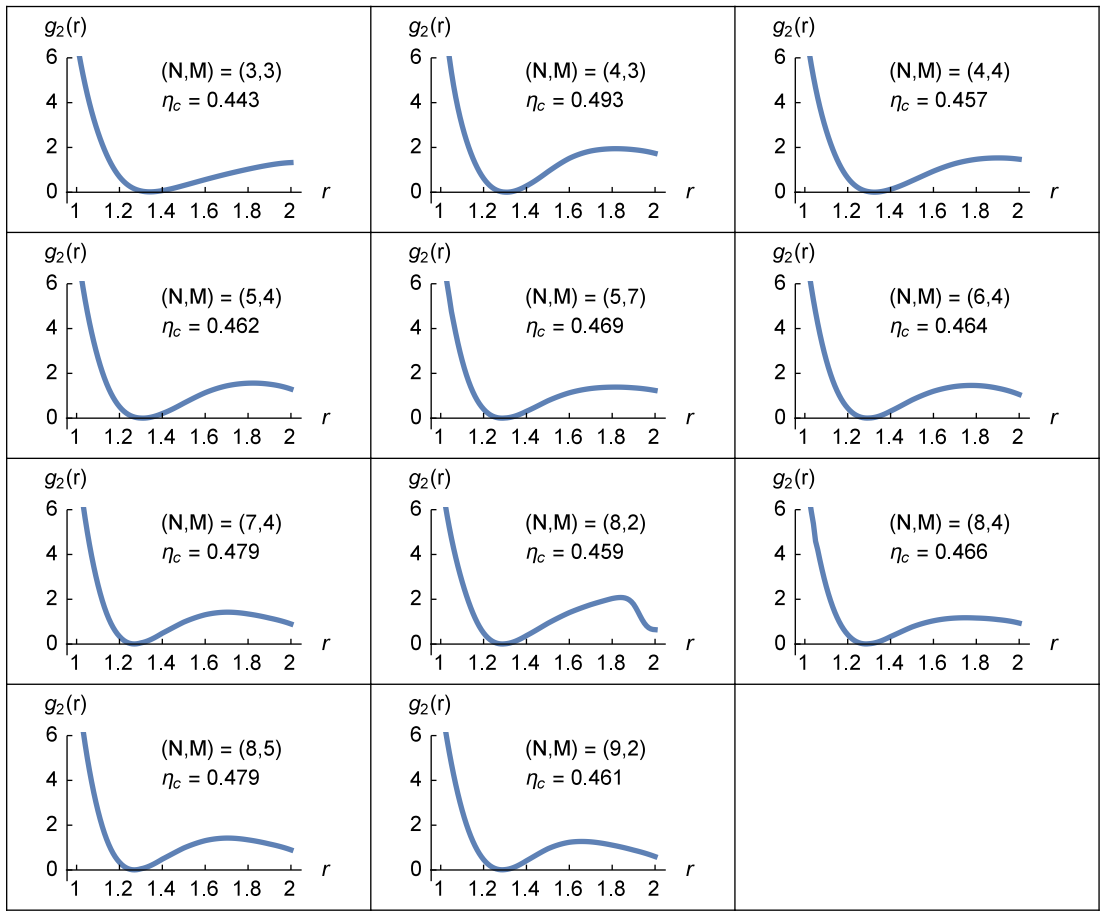}% Here is how to import EPS art
\caption{\label{fig:4}The various Pad\'e approximants of the PY pair distribution function in $d = 4$. In each case, the order of the approximation is indicated along with the resulting terminal density. The final estimate is based on the average of these different estimates. It results in $\eta_c(d=4) \simeq 0.467 \pm 0.013$.}
\end{figure*}

\noindent
The coefficients of the two polynomials are chosen in such a way, that the expansion of the ratio of the two polynomials, recovers the series 
$g_2(r) = \sum\limits_{n \ge 0} {\eta ^n}g_2^{(n)}(r)$ up to order $(N + M)$ in $\eta$. Since there are in principle many ways to choose $N$ and $M$, we mapped all the options up to order $N + M = 13$ and looked at the density for which the first zero crossing occurs. We considered only the Pad\'e approximants for which no spurious pole appears inside the interval $1 < r/R < 2$, i.e., no spontaneous divergence appears where we expect no real physical divergence to occur (this is a well-known artifact of the Pad\'e method). The various results in $d = 4$ are presented in Fig. \ref{fig:4}, and they lead to the following estimate of the largest volume fraction $\eta_c(d = 4) \simeq 0.467 \pm 0.013$. A similar analysis has been performed for $d = 6$ and $8$ and the results are summarized in Table \ref{tab:1}, as well as graphically in Fig. \ref{fig:5} (blue squares).

\begin{table}
\caption{\label{tab:1}A summary of the values $\eta_c(d)$ for dimensions in the range $3 \le d \le 9$.}
\begin{ruledtabular}
\begin{tabular}{cccccccc}
d&3&4&5&6&7&8&9\\
\hline
$\eta_c(d)$ & $0.613$ & $0.467$ & $0.367$ & $0.230$ & $0.207$ & $0.087$ & $0.112$ 
\end{tabular}
\end{ruledtabular}
\end{table}

\section{Discussion} 

The next question to consider is whether the appearance of $\eta_c(d)$ carries more physical meaning than the obvious one, namely, the inadequacy of the PY equation at high volume fraction. If indeed it carries any physical meaning it must be related to the termination of the super dense, rotation invariant steady-state, metastable phase. 
What is that termination about? The termination point should be the point where the pressure diverges.
A natural candidate is the volume fraction of random close packing (RCP).
That concept is a bit vague, however, as evident from the wide range of RCP volume fraction obtained in three dimensions
(the values of RCP volume fraction in three dimensions, obtained by numerous authors \cite{Bernal1960,Robles1998,Scott1969,Finney1970,Berryman1983,Visscher1972, Tobochnik1988} are spread between $0.6$ \cite{Visscher1972} and $0.68$ \cite{Tobochnik1988} depending on the method of derivation).
%The volume fraction at that termination point should be identified with that of the Random Close Packing density (RCP), or the alternative Maximal Random Jammed density (MRJ).
%Thus, in the following, we compare our critical volume fractions $\eta_c(d)$ with available RCP or MRJ volume fractions reported in the literature. Consider first the three dimensional case where a lot of data exists. The critical volume fraction we obtain in three dimensions is $\eta_c(3) = 0.613$.  
Due to this unsatisfactory situation, Torquato, Truskett and Debenedetti criticized the validity of the concept of RCP altogether and introduced instead the concept of MRJ \cite{Torquato2000,Torquato2010}, which is defined as the maximal random packing among all jammed configurations. The MRJ volume fraction is about $0.64$ well within the range of RCP volume fractions obtained by others. Skoge {\it et al.} \cite{Skoge2006} give the MRJ volume fraction in four, five and six dimensions and suggest also a fit for the MRJ density as a function of dimension for $3 \le d \le 6$,

\begin{equation}
\eta_{\rm{MRJ}} = \frac{c_1 + c_2 d}{2^d} ,
\label{eq:16}
\end{equation}

\noindent
where $c_1=- 2.72$ and $c_2=2.56$. 

More recently, Parisi and Zamponi \cite{Parisi2006}, using Replica theory, 
predicted a rich phase space that contains some transition densities describing 
the sphere packing in the super dense regime. 
In particular, they predicted the $J$-point, $\phi_J$,
the point where the system jams, 
and the Glass Close Packing (GCP) density, $\phi_{\rm{GCP}}$,
which is the point where the pressure diverges,
and beyond which there is no disordered packing.
Therefore, the termination point we are talking about should be compared to the GCP density, whose large $d$ dependence 
is given by \cite{Parisi2006}

\begin{equation}
\eta_{\rm{GCP}} \propto \frac{d \cdot \ln d}{2^d} .
\label{eq:17}
\end{equation}

\begin{figure}
\includegraphics[width=7cm]{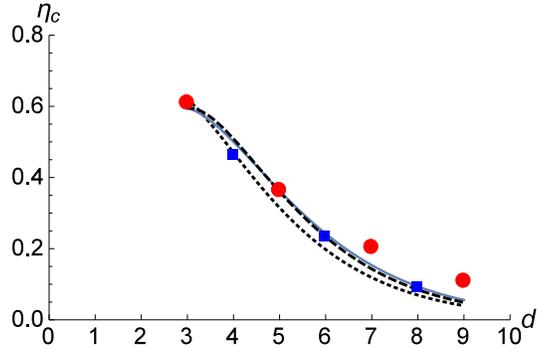}% Here is how to import EPS art
\caption{\label{fig:5} A plot of $\eta_c(d)$ for odd (red circles) and even (blue squares) dimensions in the range $3 \le d \le 9$. We also present the theoretical predictions for the MRJ density based on Eq. (\ref{eq:16}) (dotted line), our fit based on Eq. (\ref{eq:16}) (solid line) and the theoretical prediction for the GCP density based on Eq. (\ref{eq:17}) (dashed line).}
\end{figure}

\noindent
We compare these results to the terminal volume fraction for which the PY pair distribution function becomes first negative $\eta_c(d)$ in Fig. \ref{fig:5}. Note the GCP density based on Eq. (\ref{eq:17}) leaves the proportionality coefficient undetermined, and we fitted it to the data in Table \ref{tab:1} for the sake of comparison. We also tried to fit our results using the functional form given by Eq. (\ref{eq:16}), giving rise to the estimated values $\hat c_1 = -5.397$ and $\hat c_2 = 3.385$. As can be seen, the fit based on Eq. (\ref{eq:16}) (solid line) and the fit based on Eq. (\ref{eq:17}) (dashed line) are almost indistinguishable within the dimensions under discussion.

%\begin{widetext}
%\begin{equation}
%gfgggggggggggggggggg
%\end{equation}
%\end{widetext}

The statement that the terminal volume fraction identified by the Percus-Yevick theory has a physical meaning seems to be justified by the observation that $\eta_c(d)$ is close in value and behaves as a function of dimension similarly to the theoretical predictions for MRJ/GCP.  This observation is, of course, not a proof, but we can still present some supporting arguments. First, some support is gained by the one dimensional result where the PY pair distribution function remains positive for all volume fractions consistent with the fact that the real one-dimensional system of hard spheres never crosses to an RCP or MRJ state as the volume fraction is increased all the way up to $\eta = 1$, where it crystallizes.
Second, one can gain further insight into why $\eta_c(d)$ may be related to RCP and MRJ. Recall that PY provides a rotation invariant steady-state solution, which above the crystallization density describes a meta-stable disordered state. The current work shows that it ceases to exist at $\eta_c(d)$, which is consistent with the definition of RCP, namely the largest density for which a disordered phase exists. From another perspective, in a frozen system we expect each particle to sit in a cage formed by other particles. Vanishing $g_2(r)$ at a certain distance indicates the existence of such a cage. To understand that, recall that $g_2({\bf r})$ is the probability density to find a particle at $\bf r$. Therefore, if $g_2(r)$ vanishes at some distance $r_c$, no particle can cross that point, which means that the particles at $r<r_c$ are blocked within this cage.
This argument is supported by simulation results \cite{Estrada2011} in which a freezing transition is accompanied by a strong decrease in the first minimum of the pair distribution function. Although these results are associated with freezing into an ordered solid, we expect this to be also the case when the frozen state is metastable and disordered. 

%This interpretation of $\eta_c$ as the onset of caging means that within the PY approach it is indeed the most disordered density for which jamming begins (or slightly below it).

The derivation of the Percus-Yevick equation presented above highlights its underlying assumptions. 
%It does take into account the hard-core interaction and the volume constraint. 
The main ingredients are the approximate treatment of the hard sphere interaction by constraining the pair distribution function and the assumption regarding the existence of a steady-state  spherically invariant state. In that sense this approximation does not assume a fluid phase as such, and can, in principle, describe a glassy phase as well. Previous attempts to relate the RCP density and the glass transition include experimental effort using colloids \cite{Vladeeren1995}, and theoretical efforts that use the Rational Function Approximation (closely related to PY) \cite{Robles1998,Yuste1998}. However, the theoretical efforts are not based on an intrinsic criterion of the theory, but rather on an assumption of the existence of a glassy phase and a comparison to an equation of state of that phase developed by Torquato \cite{Torquato1995,Rintoul1996}. This approach has the advantage of determining the glass transition density $\eta_g$, which is not available in our approach. 
Basically, PY does not take into account the dynamics and hence it is clear that one would need further input if it is to be used to describe the glass transition. We already explained that PY cannot capture any nontrivial higher order correlation or response functions, and definitely not time-dependent quantities, such as $\chi_4$, which are extensively used in the glass community \cite{Toninelli2005,Berthier2011a,Berthier2011b} to characterize the glass transition. This observation motivates future effort to take into account the dynamics (e.g. in the Fokker-Planck Eq. (\ref{eq:4}) along the lines of Ref. \cite{Schwartz1999}) as well as go beyond the present approximation (via constraining the triplet distribution function and using the self-consistent expansion), to gain more insight into the glassy state. Actually, this may be particularly useful since previtrification arising from ergodicity breaking \cite{Brito2009} is already present way below the glass transition in the supercooled phase.

\section{Summary and future perspective} 

To summarize, in this paper we show that unlike the common lore, the Percus-Yevick theory for monodisperse hard spheres provides an intrinsic indication for its limitation in the regime of high densities. More specifically, the positivity of the pair-correlation function $g_2({\bf r})$ in $d$ dimensions is violated at a certain volume fraction which we denote $\eta_c(d)$, and thus beyond it the PY theory is no longer consistent. It turns out that this phenomenon occurs in all dimensions in the range $3 \le d \le 9$, suggesting that it should hold also beyond. A comparison of $\eta_c(d)$ to the various results and predictions for the RCP, GCP or MRJ volume fractions shows that they are close and behave similarly as a function of dimension. This observation suggests that the terminal volume fraction in the PY theory actually indicates the largest density for which a spherically invariant state can exist, even if the solid phase is already preferred at this point, and without being able to distinguish a fluid from a glass. In that sense, the PY theory can provide a simple approximation or indication for the existence of a termination density. Further investment using more elaborate methods such as replica theory could then reveal the full picture, which may contain glassy phases.

We hope this work will motivate other researchers to check this phenomenon in many other systems described by a Percus-Yevick theory. A few examples are - hard spheres in curved space \cite{Lishchuk2006,Lishchuk2009,Sausset2009,Tarjus2011} and hard spheres experiencing more complicated interactions such as sticky hard spheres \cite{Sengers2000,Baxter1968b} or  square-well fluids \cite{Mulero2008}. Other important direction are systems composed of polydisperse or mixture of hard spheres \cite{Hansen2006,Sengers2000,Mulero2008,Brouwers2006,Lebowitz1964,Gray2011,Yuste1998}, various charged hard sphere fluids \cite{Sengers2000} such as the hard sphere Yukawa fluid \cite{Chialvo1996}, ionic liquids \cite{Hansen2006,Mulero2008}, polarizable fluids \cite{Gray2011}, and even fluids of non-spherical shapes such as ellipsoids \cite{Letz1999,Khordad2009}, spherocylinders \cite{Martinez2003}, and chain-like molecules \cite{Sengers2000,Mulero2008}.

%\begin{acknowledgments}
%We wish to acknowledge the support of ???
%\end{acknowledgments}

\nocite{*}
%\bibliography{aipsamp}% Produces the bibliography via BibTeX.

%\end{document}

\newpage

%\section{Supplemental Material to the paper ``Random Close Packing and the Hard Sphere Percus-Yevick Theory" by Eytan Katzav, Ruslan Berdichevsky and Moshe Schwartz}

%\begin{appendices}

\appendix
\renewcommand{\theequation}{A\arabic{equation}}
\renewcommand\thefigure{\thesection\arabic{figure}}    
\setcounter{figure}{0}   
%\section{Appendixes}
%\verb+\appendix+
\section{The Percus-Yevick approximation in terms of the direct correlation function}

The Percus-Yevick approximation derived in section II and summarized by Eqs. (\ref{eq:9}) and (\ref{eq:10}) is usually written in terms of the direct correlation function $c(r)$, which is related to the Lagrange multiplier introduced in Eq. (\ref{eq:7}) by $c({\mathbf{r}}) = -\lambda ({\mathbf{r}})/\bar \rho $. The direct correlation,$c({\mathbf{r}})$ is determined by the so-called Ornstein-Zernike equation,

\begin{equation}
h({\mathbf{r}}) = c({\mathbf{r}}) + \bar \rho \int\limits_0^\infty  {h({{\mathbf{r'}}} )c({\left| {{\mathbf{r - r'}}} \right|} )d{\mathbf{r'}}} ,
\label{eq:18}
\end{equation}

\noindent
where $h({\mathbf{r}}) = g_2({\mathbf{r}}) - 1$ is called the total correlation function. Note that this equation is equivalent to Eq. (\ref{eq:10}) by using Eqs. (\ref{eq:8}) and (\ref{eq:9}). Here and in the following, we take $r$ in units of $R$ , the diameter of the hypersphere to be unity, and thus In $d$ dimensions we have for the volume fraction $\eta$

\begin{equation}
\eta  = \bar \rho V_d \left(\frac{R}{2}\right) = \left(\frac{\pi}{4}\right)^{d/2}
\frac{\bar \rho R^d}{\Gamma \left( {{\textstyle{{d + 2} \over 2}}} \right)} ,
\label{eq:19}
\end{equation}

\noindent
where $V_d(r)$ is the volume of a $d$-dimensional hypersphere of radius $r$, and $\Gamma(x)$ is the Euler gamma function. This equation generalizes Eq. (\ref{eq:13}) to any dimension.

In odd dimensions, a highly non trivial result \cite{Leutheusser1984,Rohrmann2007,Robles2007,Robles2004} is that the direct correlation function $c(r)$ within the PY approximation turns out to be a polynomial of degree $d$, namely \\
$c(r) = \theta \left( 1 - r/R \right)\sum\limits_{i = 0}^d {c_i (\eta) (r/R)^i}$, where $\theta (x)$ is the Heaviside function. Therefore, obtaining the corresponding Fourier components $\tilde c(k)$ is a straightforward analytical calculation

\begin{equation}
\tilde c(k) = (2\pi)^{d/2} k^{-\frac{d-2}{2}}
\sum\limits_{i = 0}^d c_i (\eta)\int\limits_0^1 u^{i + d/2} J_{\frac{d-2}{2}}(ku) du ,
\label{eq:20}
\end{equation}

\noindent
where $J_\nu (x)$ is the Bessel function order $\nu$.
From this the structure factor is obtained via $S(k) = 1/\left( {1 - \bar \rho \tilde c(k)} \right)$, and the radial distribution function $g_2(r)$ can be obtained using Eq. (\ref{eq:10}). Note that in the dimensions discussed in the appendices, namely for $d>3$, there is no direct analytical representation of $g_2(r)$ as the one available in three dimensions \cite{Smith1970,Kelly2016}, and therefore there is no alternative to performing a numerical inverse Fourier transform.

\renewcommand{\theequation}{B\arabic{equation}}
\setcounter{equation}{0}
\section{The pair distribution functions in five, seven and nine dimensions}

In this part we show the function $g_2(r)$ in five (Figs. B1 and B2), seven (Figs. B3 and B4) and nine (Figs. B5 and B6) dimensions similarly to what has been presented in the text for one and three dimensions. As can be seen in all cases there is a critical volume fraction 
$\eta_c(d)$ at which $g_2(r)$ starts to develop a negative part, which marks the termination density of applicability of the PY theory.

\begin{figure}
\includegraphics[width=7cm]{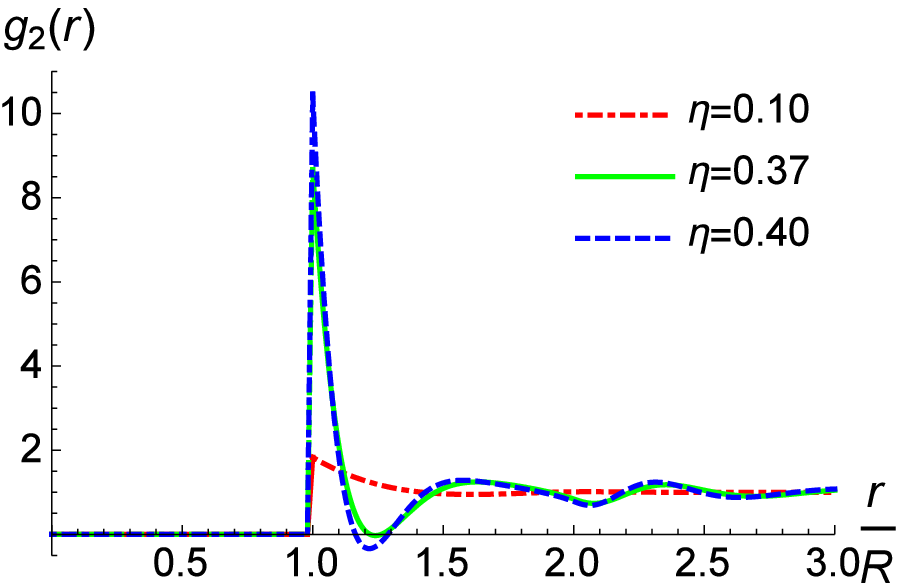}
\caption{\label{fig:B1} The five dimensional pair distribution functions for low, intermediate and high volume fractions.}
\end{figure}

\begin{figure}
\includegraphics[width=7cm]{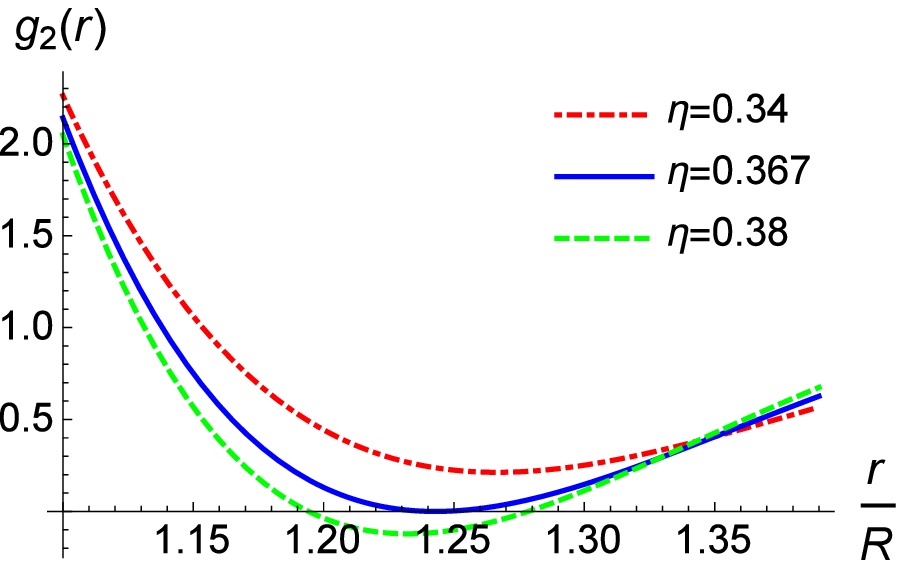}
\caption{\label{fig:B2} The five dimensional PY pair distribution function: a zoom into the region where it becomes negative. As can be seen from the figure $\eta_c (5) \simeq 0.367$.}
\end{figure}

\begin{figure}
\includegraphics[width=7cm]{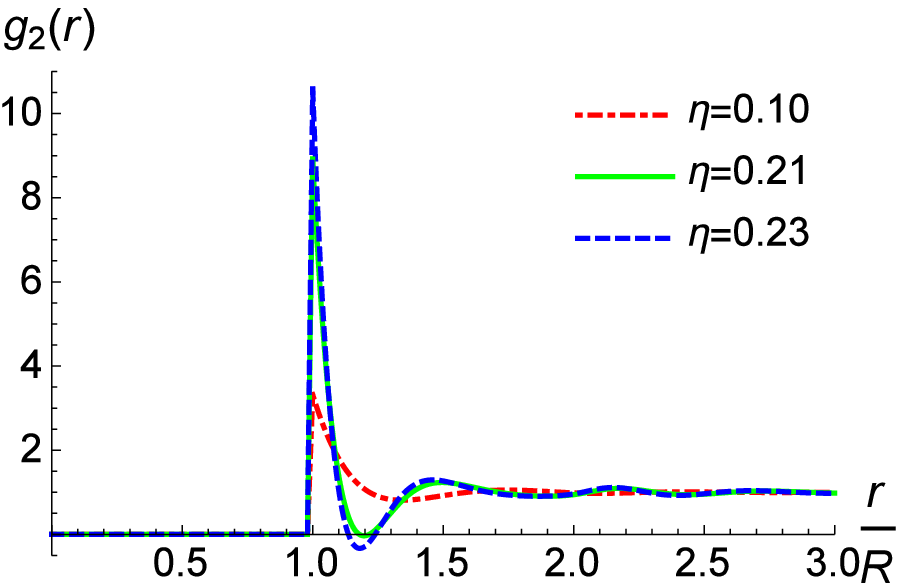}
\caption{\label{fig:B3} The seven dimensional pair distribution functions for low, intermediate and high volume fractions.}
\end{figure}

\begin{figure}
\includegraphics[width=7cm]{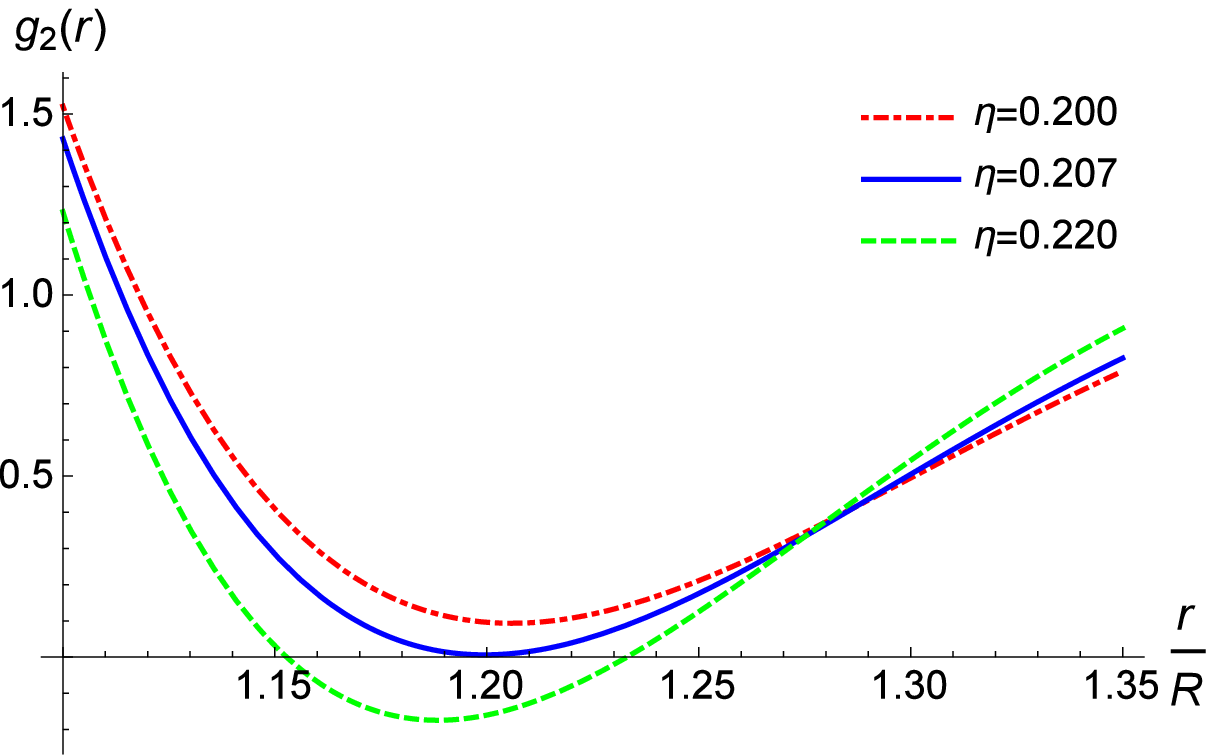}
\caption{\label{fig:B4} The seven dimensional PY pair distribution function: a zoom into the region where it becomes negative. As can be seen from the figure $\eta_c(7) = 0.207$.}
\end{figure}

\begin{figure}
\includegraphics[width=7cm]{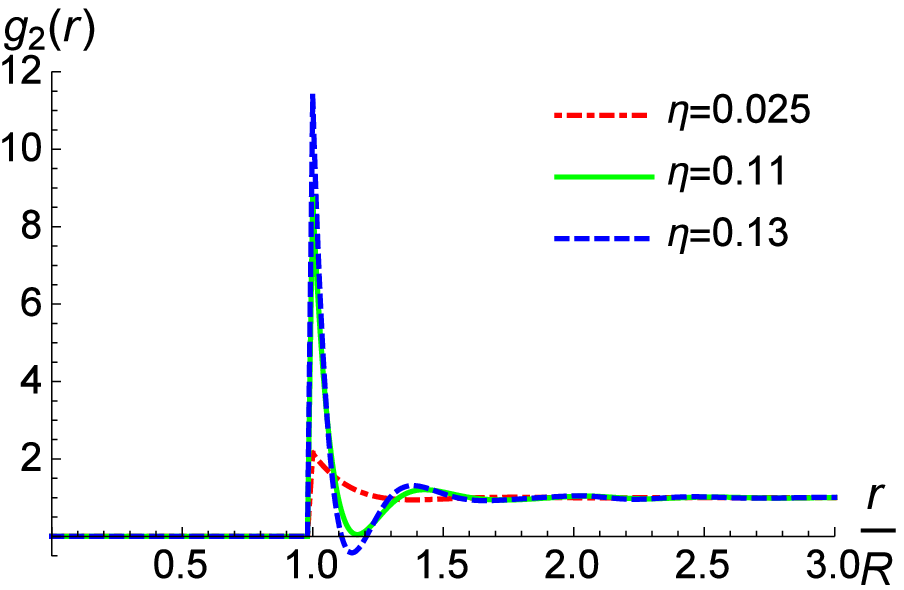}
\caption{\label{fig:B5} The nine dimensional pair distribution functions for low, intermediate and high volume fractions.}
\end{figure}

\begin{figure}
\includegraphics[width=7cm]{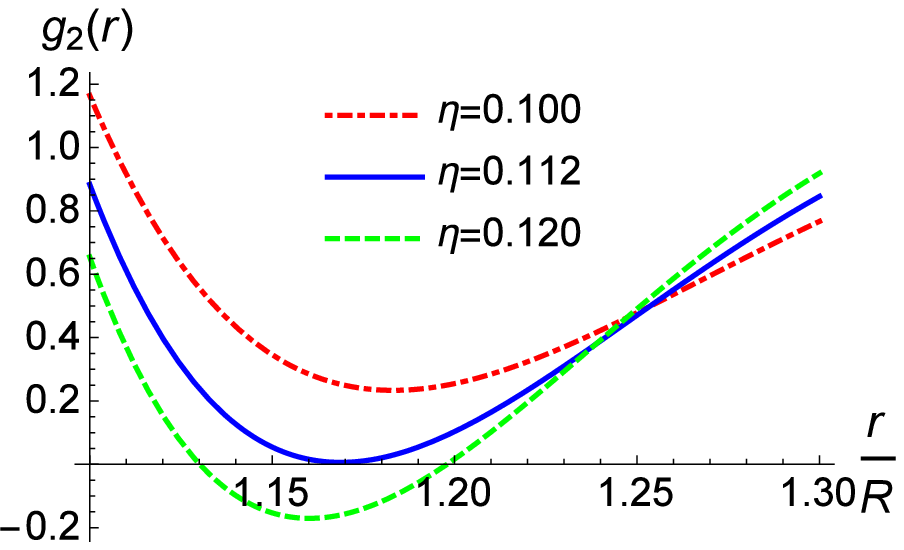}
\caption{\label{fig:B6} The nine dimensional PY pair distribution function: a zoom into the region where it becomes negative. As can be seen in the figure $\eta_c(9) = 0.112$.}
\end{figure}

\end{document}